\documentclass[conference]{IEEEtran}

\IEEEoverridecommandlockouts

\usepackage{cite}
\usepackage{amsmath,amssymb,amsfonts}
\usepackage{algorithmic}
\usepackage{graphicx}
\usepackage{textcomp}
\usepackage{xr}
\usepackage[table]{xcolor}

\def\BibTeX{{\rm B\kern-.05em{\sc i\kern-.025em b}\kern-.08em
		T\kern-.1667em\lower.7ex\hbox{E}\kern-.125emX}}

\usepackage{tikz}
\usetikzlibrary{backgrounds}
\usetikzlibrary{spy}
\usetikzlibrary{calc,arrows.meta}
\usepackage{pgfplots}
\usetikzlibrary{plotmarks}
\usetikzlibrary{arrows.meta}
\usetikzlibrary{positioning,fit}
\usetikzlibrary{intersections}
\usetikzlibrary{patterns}
\usetikzlibrary{decorations.shapes}

\usepgfplotslibrary{patchplots}
\usepgfplotslibrary{colormaps}

\usepackage{cancel}

\usepackage{pdfpages} %

\usepackage{multirow}
\usepackage{makecell}

\usepackage{algorithmic}
\usepackage[ruled,vlined]{algorithm2e}
\SetKwRepeat{Do}{do}{while}

\newcommand{\exportFigures}{true}
\newcommand{\exportFiguresAsPNG}{true}

\ifthenelse{\equal{\exportFigures}{true}}
{
	\usepgfplotslibrary{external}
	\tikzexternalize[prefix=compiled_tikz_figures/,optimize command away=\includepdf]
	\ifthenelse{\equal{\exportFiguresAsPNG}{true}}
	{
		\tikzset
		{   png export/.style={
				external/system call={
					pdflatex \tikzexternalcheckshellescape -halt-on-error --extra-mem-top=10000000 -interaction=batchmode -jobname "\image" "\texsource" && pdftops -eps "\image.pdf" && convert -density 700 -transparent white "\image.pdf" "\image.png"
		}}}
		\tikzset{png export}
	}
	{}
}
{}

\usepackage{url}

\usepackage{bm}
\usepackage[caption=false,font=footnotesize]{subfig}
\usepackage[normalem]{ulem}

\usepackage{mathtools, cuted}

\usepackage{acronym}

\usepackage{booktabs}
		
\usepackage{pdfpages}

\usepackage[latin1]{inputenc}
\usepackage{tikz}
\usetikzlibrary{shapes,arrows}
\usetikzlibrary{arrows.meta}
\usetikzlibrary{positioning}

\usepackage{ellipsis}

\usetikzlibrary{calc}
\usetikzlibrary{decorations.pathreplacing,decorations.markings,shapes.geometric}
\usetikzlibrary{decorations.pathmorphing}
\usetikzlibrary{fit}
\usetikzlibrary{pgfplots.groupplots}

\usetikzlibrary{calc,arrows.meta}

\usetikzlibrary{backgrounds} %

\definecolor{green(pigment)}{rgb}{0.0, 0.65, 0.31}
\definecolor{frenchblue}{rgb}{0.0, 0.45, 0.73} 
\definecolor{mediumcandyapplered}{rgb}{0.89, 0.02, 0.17}

\usepackage[most]{tcolorbox}
\tcbuselibrary{breakable}
\tcbset{every box/.style={enhanced,breakable}}
\tcbset{colframe=black,colback=red!10,enhanced,breakable,sharp corners}

\usepackage{enumitem} %

\usepackage{notation}

\usepackage{adjustbox}

\definecolor{alex}{RGB}{51,183,150}
\definecolor{erik}{RGB}{235,134,52}

\newcommand{\ticked}{$\text{\rlap{$\checkmark$}}\square$}
\newcommand{\unticked}{{$\square$}}
\newcommand{\tick}[1]{\ifthenelse{#1=1}{\ticked}{\unticked}}

\newcommand{\rmv}{\hspace*{-.3mm}}
\renewcommand{\Re}[1]{\ensuremath{\text{Re}\!\left[#1\right]}}%
\renewcommand{\Im}[1]{\ensuremath{\text{Im}\!\left[#1\right]}}%

\newcommand{\norm}[2]{\ensuremath{\lVert #1 \rVert^{#2}}}%

\newcommand{\minus}{\rmv - \rmv}

\newcommand{\s}{\hspace*{0.5pt}}

\newcommand{\pe}{p_{\text{E}}(u_n^{(j)}\rmv\rmv\rmv,q_n^{(j)})}

\newcommand{\pdrv}{p_{\text{D}}(\rv{u}_n^{(j)})}

\providecommand{\norm}[1]{\lVert#1\rVert}

\newcommand{\zf}{\ensuremath{{z^{(j)}_\text{f}}_{\hspace{-2.6mm} i,n}}}

\newcommand{\zd}{\ensuremath{{z_\mathrm{d}^{(j)}}_{\rmv\rmv\rmv\rmv\rmv\rmv\rmv  m,n}}}
\newcommand{\zu}{\ensuremath{{z_\mathrm{u}^{(j)}}_{\rmv\rmv\rmv\rmv\rmv\rmv\rmv m,n}}}
\newcommand{\zdr}{\ensuremath{{\rv{z}_\mathrm{d}^{(j)}}_{\rmv\rmv\rmv\rmv\rmv\rmv\rmv  m,n}}}
\newcommand{\zur}{\ensuremath{{\rv{z}_\mathrm{u}^{(j)}}_{\rmv\rmv\rmv\rmv\rmv\rmv\rmv m,n}}}

\mathchardef\Re="023C
\mathchardef\Im="023D

\newlength{\figureheight}
\newlength{\figurewidth}
\graphicspath{{./figures/}}

\allowdisplaybreaks

\definecolor{mycolor01}{rgb}{0.00000,0.00000,1.00000}
\definecolor{mycolor02}{rgb}{0.133,0.545,0.133}
\definecolor{mycolor03}{rgb}{0.50000,0.00000,0.50000}
\definecolor{mycolor05}{rgb}{1.00000,0.83984,0.00000}
\definecolor{mycolor04}{rgb}{0.92969,0.50781,0.92969}
\definecolor{mycolor06}{rgb}{1.00000,0.64453,0.00000}
\definecolor{mycolor07}{rgb}{0.50000,0.50000,0.50000}
\definecolor{mycolor08}{rgb}{1.00000,0.00000,0.00000}
\definecolor{mycolor09}{rgb}{0.2510 ,0.8784, 0.8157}
\definecolor{mycolor10}{rgb}{0.54297,0.00000,0.00000}
\definecolor{mycolor11}{rgb}{0.6445, 0.1641,0.1641}
\definecolor{mycolor12}{rgb}{1, 0, 1}

\pgfdeclarelayer{back1}
\pgfdeclarelayer{lay1}
\pgfdeclarelayer{back2}
\pgfdeclarelayer{lay2}
\pgfsetlayers{back2,lay2,back1,main,lay1}

\makeatletter

\tikzset{
	nomorepostactions/.code={\let\tikz@postactions=\pgfutil@empty},
	decmark/.style 2 args={decoration={markings,
			mark= between positions 0 and 1 step (1/6)*\pgfdecoratedpathlength with{%
				\tikzset{#2,every mark}\tikz@options
				\pgftransformresetnontranslations
				\pgfuseplotmark{#1}%
			},  
		},
		postaction={decorate},
		/pgfplots/legend image post style={
			mark=#1, mark options={#2}, every path/.append style={nomorepostactions}
		},
	},
	markbeginend/.style 2 args={decoration={markings,
			mark= between positions 0 and 1 step (1)*\pgfdecoratedpathlength with{%
				\tikzset{#2,every mark}\tikz@options
				\pgfuseplotmark{#1}%
			},  
		},
		postaction={decorate},
		/pgfplots/legend image post style={
			mark=#1,mark options={#2},every path/.append style={nomorepostactions}
		},
	},
	markend/.style 2 args={decoration={markings,
			mark= at position \pgfdecoratedpathlength with{%
				\tikzset{#2,every mark}\tikz@options
				\pgfuseplotmark{#1}%
			},  
		},
		postaction={decorate},
		/pgfplots/legend image post style={
			mark=#1,mark options={#2},every path/.append style={nomorepostactions}
		},
	},
	posmark/.style 2 args={decoration={markings,
			mark= at position #2 with{%
				\tikzset{solid,every mark}\tikz@options
				\pgftransformresetnontranslations
				\pgfuseplotmark{#1}%
			},  
		},
		postaction={decorate},
		/pgfplots/legend image post style={
			mark=#1,mark options={solid},every path/.append style={nomorepostactions}
		},
	},
}

\makeatother

\pgfplotsset{
	resultStyle1/.style={mark=none, line width=0.5pt, mycolor01, decmark={oplus}{solid}},
	resultStyle2/.style={mark=none, line width=0.5pt, mycolor02, decmark={triangle}{solid}},%
resultStyle3/.style={mark=none ,line width=0.5pt, mycolor03, decmark={+}{solid}},
resultStyle4/.style={mark=none, line width=0.5pt, mycolor06, decmark={star}{solid}},
resultStyle5/.style={mark=none, line width=0.5pt, mycolor08, decmark={o}{solid}},
resultStyle6/.style={mark=none, line width=0.5pt, mycolor05, decmark={square}{solid}}, 
resultStyle7/.style={mark=none, line width=0.5pt, mycolor09, decmark={diamond}{solid}}, 
resultStyle8/.style={mark=none, line width=0.5pt, mycolor11, decmark={otimes}{solid}}, 
resultStyle9/.style={mark=none, line width=0.5pt, mycolor12, decmark={x}{solid}}, 
resultStyleBase/.style={mark=none, line width=0.5pt,}, 
compareStyle1/.style={mark=none, line width=0.5pt, mycolor01},
compareStyle2/.style={mark=none, line width=0.5pt, mycolor02},%
compareStyle3/.style={mark=none ,line width=0.5pt, mycolor03},
compareStyle4/.style={mark=none, line width=0.5pt, mycolor06},
compareStyle5/.style={mark=none, line width=0.5pt, mycolor08},
compareStyle6/.style={mark=none, line width=0.5pt, mycolor05}, 
compareStyle7/.style={mark=none, line width=0.5pt, mycolor09}, 
compareStyle8/.style={mark=none, line width=0.5pt, mycolor11}, 
compareStyle9/.style={mark=none, line width=0.5pt, mycolor12}, 
}

\pgfplotsset{
compat=newest,
simple style group/.style={
label style={font=\scriptsize},
legend style={font=\scriptsize},
tick label style={font=\scriptsize},
nodes near coords style={font=\scriptsize},
title style={font=\scriptsize},
scale only axis,
grid style={dotted},
mark options={solid}, %
},
simple style/.style={
label style={font=\scriptsize},
legend style={font=\scriptsize},
tick label style={font=\scriptsize},
nodes near coords style={font=\scriptsize},
title style={font=\scriptsize},
width=\figurewidth,
height=\figureheight,
at={(0\figurewidth,0\figureheight)},
scale only axis,
grid style={dotted},
mark options={solid}, %
},
base style/.style={
label style={font=\scriptsize},
legend style={font=\scriptsize},
tick label style={font=\scriptsize},
nodes near coords style={font=\scriptsize},
title style={font=\scriptsize},
width=\figurewidth,
height=\figureheight,
at={(0\figurewidth,0\figureheight)},
scale only axis,
cycle list={
	{mark=none, line width=0.5pt, mycolor01, solid},
	{mark=none, line width=0.5pt, mycolor02, dash dot},
	{mark=none ,line width=0.5pt, mycolor03, densely dashed},
	{mark=none, line width=0.5pt, mycolor04, dash dot dot},
	{mark=x   , line width=0.5pt, mycolor05},
	{mark=.   , line width=0.7pt, mycolor06}, 
	{mark=square,only marks, mark size = 0.8pt, mycolor07,
		mark options = {line width = 0.4pt}},
	{mark=x,     only marks, mark size = 1.3pt, mycolor08,
		mark options = {line width = 0.4pt}},
	{mark=o,     only marks, mark size = 0.8pt, mycolor09,
		mark options = {line width = 0.4pt}},
	{mark=o, mycolor10},
},
grid style={dotted},
xmajorgrids,
ymajorgrids,
mark options={solid}, %
},
base style group/.style={
	label style={font=\scriptsize},
	legend style={font=\scriptsize},
	tick label style={font=\scriptsize},
	nodes near coords style={font=\scriptsize},
	title style={font=\scriptsize},
	scale only axis,
	grid style={dotted},
	xmajorgrids,
	ymajorgrids,
	mark options={solid}, %
},
std graph style new/.style={
xlabel style={yshift=1mm},
ylabel style={yshift=-1.5mm},
yticklabel style={xshift=1mm},
},
color lines style/.style={
cycle list={
	{mark=none, mycolor01, decmark={oplus}{solid} },
	{mark=none, mycolor02, decmark={+}{solid} }, 
	{mark=none, mycolor03, decmark={triangle}{solid} }, 
	{mark=none, mycolor04, decmark={star}{solid} }, 
	{mark=none, mycolor05, decmark={o}{solid} },
	{mark=none, mycolor06, decmark={square}{solid} },
},
},
meas graph style/.style={
xlabel style={yshift=1mm},
ylabel style={yshift=-1mm},
xmajorgrids,
ymajorgrids,
mark repeat = 1,
mark phase = 0,
cycle list={
	{color=black, only marks, mark=*, mark size=0.5pt, mark options={solid, black}},
	{color=red, only marks, mark=*, mark size=0.1pt, line width=0.25pt},
},
ylabel={},
}, 
ci graph style/.style={
xlabel style={yshift=1mm},
ylabel style={yshift=-1.5mm},
yticklabel style={xshift=1mm},
mark repeat = 1,
mark phase = 0,
ymin=1e-3,
ymax=100,
ytick = {100, 50, 10, 1, 0.1, 0.01, 1e-3, 1e-4},
yticklabels = {$0$, $50$, $90$, $99$, $99.9$, $99.99$, $99.999$, $99.9999$},
y dir=reverse,
},     
bp coeff style/.style={
scale only axis=true,
width=0.225*.9\linewidth,
height=0.225*.9\linewidth,
scale only axis,
xmin=-4.000,
xmax=4.000,
xlabel={$\ell${\color{white}$\aod$}},
ticklabel style={font=\footnotesize},
ymin=0.000, ymax=0.9,
ylabel={$c_\ell$},
xlabel style={font=\footnotesize},
ylabel style={font=\footnotesize},
major tick length=2pt%
},
bp graph style/.style={        
scale only axis=true,
width=0.35*1.1\linewidth,
height=0.225*.9\linewidth,
scale only axis,
xmin=-3.14, xmax=3.14,
xlabel={$\aod${\color{white}$\ell$}},
ticklabel style={font=\footnotesize},
xtick={-3.14,-1.57,0.0,1.57,3.14},
xticklabels={$-\pi$,$-\tfrac{\pi}{2}$,$0$,$\tfrac{\pi}{2}$,$\pi$},
ymin=0.000, ymax=3,
ylabel={Beampattern},
xlabel style={font=\footnotesize}, ylabel style={font=\footnotesize},
major tick length=2pt
},
peb graph style/.style={        
width=0.66\linewidth,%
scale only axis,
point meta min=-2.583,
point meta max=-0.300,
axis on top,
xmin=0.000,
xmax=12.000,
xlabel={x in meter},
y dir=reverse,
ymin=0.000,
ymax=8.000,
ylabel={y in meter},
ytick={7.0,6.0,...,0.0},
xtick={0.0,1.0,...,12.0},
yticklabels={$1$,$2$,$3$,$4$,$5$,$6$,$7$,$8$},
xlabel style={font=\scriptsize,yshift=0.125cm},
ylabel style={font=\scriptsize,yshift=-0.125cm},
ticklabel style={font=\scriptsize},
unit vector ratio*=1 1 1,
yticklabel pos=left,
major tick length=2pt,
colormap={mymap}{[1pt] rgb(0pt)=(1,1,1); rgb(1pt)=(0.858903,0.984776,0.839302); rgb(2pt)=(0.777958,0.94143,0.649487); rgb(3pt)=(0.755504,0.864264,0.463393); rgb(4pt)=(0.777509,0.754439,0.310168); rgb(5pt)=(0.820314,0.619497,0.21003); rgb(6pt)=(0.854796,0.471879,0.170327); rgb(7pt)=(0.851327,0.326629,0.183322); rgb(8pt)=(0.784671,0.198575,0.225774); rgb(9pt)=(0.637629,0.0993149,0.259577); rgb(10pt)=(0.400067,0.0343393,0.229819); rgb(11pt)=(0,0,0)},
colorbar style={ylabel={Position Error Bound in centimeter (logscale)}, ytick={-0.4,-0.82,...,-2.92}, yticklabels={$39.8$, $15.1$, $5.8$, $2.2$, $0.8$, $0.3$},ylabel style={yshift=0.5mm,font=\scriptsize,scale=0.8},width=2.0mm,xshift=-4.25mm,ticklabel style={font=\scriptsize},major tick length=0pt}, %
colormap access=piecewise constant
},
peb ellipses/.style={color=white, line width=0.4pt, forget plot}
}

\tikzset{naming/.style={align=center,font=\small}}
\tikzset{antenna/.style={insert path={-- coordinate (ant#1) ++(0,0.25) -- +(135:0.25) + (0,0) -- +(45:0.25)}}}
\tikzset{station/.style={naming,draw,shape=dart,shape border rotate=90, minimum width=10mm, minimum height=10mm,outer sep=0pt,inner sep=3pt}}
\tikzset{mobile/.style={naming,draw,shape=rectangle,minimum width=12mm,minimum height=6mm, outer sep=0pt,inner sep=3pt}}
\tikzset{radiation/.style={{decorate,decoration={expanding waves,angle=90,segment length=4pt}}}}

\tikzset{
  pobl/.style={
    inner sep=0pt, outer sep=0pt, fill=#1,
  },
  pobl gron/.style n args={2}{
    pobl=#1, rounded corners=#2,
  },
  pics/person/.style n args={3}{
    code={
      \node (-corff) [pobl=#1, minimum width=.25*#2, minimum height=.375*#2, rotate=#3, pic actions] {};
      \node (-pen) [minimum width=.3*#2, circle, pobl=#1, outer sep=.01*#2, anchor=south, rotate=#3, pic actions] at (-corff.north) {};
      \node (-coes dde) [pobl gron={#1}{1pt}, anchor=north west, minimum width=.12125*#2, minimum height=.25*#2, rotate=#3, pic actions] at (-corff.south west) {};
      \node [pobl=#1, anchor=north, minimum width=.12125*#2, minimum height=.15*#2, rotate=#3, pic actions] at (-coes dde.north) {};
      \node (-coes chwith) [pobl gron={#1}{1pt}, anchor=north east, minimum width=.12125*#2, minimum height=.25*#2, rotate=#3, pic actions] at (-corff.south east) {};
      \node [pobl=#1, anchor=north, minimum width=.12125*#2, minimum height=.15*#2, rotate=#3, pic actions] at (-coes chwith.north) {};
      \node (-braich dde) [pobl gron={#1}{.75pt}, minimum width=.075*#2, minimum height=.325*#2, outer sep=.0064*#2, anchor=north west, rotate=#3, pic actions] at (-corff.north east)  {};
      \node [pobl=#1, minimum width=.05*#2, minimum height=.2*#2, outer sep=.0064*#2, anchor=north west, rotate=#3, pic actions] at (-corff.north east) {};
      \node (-braich chwith) [pobl gron={#1}{.75pt}, minimum width=.075*#2, minimum height=.325*#2, outer sep=.0064*#2, anchor=north east, rotate=#3, pic actions] at (-corff.north west) {};
      \node [pobl=#1, minimum width=.0375*#2, minimum height=.2*#2, outer sep=.0064*#2, anchor=north east, rotate=#3, pic actions] at (-corff.north west) {};
      \node (-fit person) [fit={(-pen.north) (-braich dde.east) (-coes chwith.south) (-braich chwith.west)}] {};
    },
  },
  pics/SBS/.style={code={
      \begin{scope}[local bounding box=#1]
      \fill [pic actions/.try] (-1,0) -- (-1/2,3) -- (1/2, 3) -- (1,0) -- cycle;
      \fill [pic actions/.try] (-1/16,2) rectangle (1/16,4);
      \fill [pic actions/.try] (0,4) circle [radius=1/4];
      \foreach \i in {-1,1}
        \fill [shift=(90:4), xscale=\i]
          \foreach \r in {1,3/2,2}{
            (-45:\r) arc (-45:45:\r) -- (45:\r-1/10)
            arc(45:-45:\r-1/10) -- cycle
          };
       \end{scope}
  }},
}

\begin{document}

\title{A Neural-enhanced Factor Graph-based Algorithm for Robust Positioning in Obstructed LOS Situations}

\author{Alexander Venus$^{1,2}$, Erik Leitinger$^{1,2}$, Stefan Tertinek$^{3}$, %
 and  Klaus Witrisal$^{1,2}$
\thanks{The financial support by the Christian Doppler Research Association, the Austrian Federal Ministry for Digital and Economic Affairs and the National Foundation for Research, Technology and Development is gratefully acknowledged.}

\\
\small{{$^1$Graz University of Technology, Austria}, {$^3$NXP Semiconductors, Austria},
}\\
\small{{$^2$Christian Doppler Laboratory for Location-aware Electronic Systems}}\\[-2mm]
}

\maketitle
\frenchspacing

\renewcommand{\baselinestretch}{0.98}\small\normalsize %

\begin{abstract}
	This paper presents a neural-enhanced probabilistic model and corresponding factor graph-based sum-product algorithm for robust localization and tracking in multipath-prone environments. 
The introduced hybrid probabilistic model consists of physics-based and data-driven measurement models %
capturing the information contained in both, the \ac{los} component as well as in \aclp{mpc} (\acs{nlos} components). %
The physics-based and data-driven models are embedded in a joint Bayesian framework allowing to derive from first principles a factor graph-based algorithm that fuses the information of these models. 
The proposed algorithm uses radio signal measurements from multiple base stations to robustly estimate the mobile agent's position together with all model parameters. %
It provides high localization accuracy by exploiting the position-related information of the LOS component via the physics-based model and robustness by exploiting the geometric imprint of \aclp{mpc} independent of the propagation channel via the data-driven model. 
In a challenging numerical experiment involving obstructed LOS\acused{olos} situations to all anchors, we show that the proposed sequential algorithm significantly outperforms state-of-the-art methods and attains the \acl{pcrlb} even with training data limited to local regions.

\end{abstract}

\acresetall %


\IEEEpeerreviewmaketitle

\section{Introduction}\label{sec:introduction}

Localization of mobile agents using radio signals is still a challenging task in environments such as indoor or urban territories\cite{WitrisalSPM2016Copy,Mendrzik2019,WangShenTWC2020}. These environments are characterized by strong multipath propagation and frequent obstructed line-of-sight (OLOS)\acused{olos} situations, which can prevent the correct extraction of information contained in the \ac{los} component\footnote{Throughout this paper we use the terms ``\ac{mpc}'' and ``NLOS components'' interchangeably to refer to all received signal components  except the \ac{los} component. Also, we use the term ``\acl{olos}'' (\acs{olos}), to refer to situations, where the LOS component is blocked or cannot be detected. We distinguish between partial and full OLOS situations, where the LOS component of some or all anchors is unavailable.}. %
There exist many safety- and security-critical applications, such as autonomous driving \cite{Karlsson2017}, medical services \cite{KoEMBMag2010}, or keyless entry systems \cite{Kalyanaraman2020}, where robustness of the position estimate (i.e, no lost tracks)  %
is of critical importance.

\subsection{State-of-the-Art Methods} \label{sec:sota}

New localization and tracking approaches within the context of 6G localization \cite{ContiCommunMag2021} take advantage of large measurement apertures as \ac{uwb} systems \cite{DardariProcIEEE2009, TaponeccoTWC2011} or mmWave systems \cite{RusekSPM2013}, which allow the received radio signal to be resolved into a superposition of a finite number of specular \acfp{mpc} \cite{RichterPhD2005, DardariProcIEEE2009,ShenTIT2010,AdityaProc2018}. Such novel approaches try to mitigate the effect of multipath propagation \cite{GiffordTSP2022} and \ac{olos} situations \cite{WymeerschIEEE2012, AdityaProc2018} or even take advantage of \acp{mpc} by exploiting inherent position information, turning multipath from impairment to an asset \cite{WitrisalSPM2016Copy,GentnerTWC2016, ShahmansooriTWC2018}.
\begin{figure}[t]
	\vspace{-1mm}
	\centering
	\tikzsetnextfilename{eye_catcher}
	\hspace*{-1mm}\scalebox{0.95}{\includegraphics{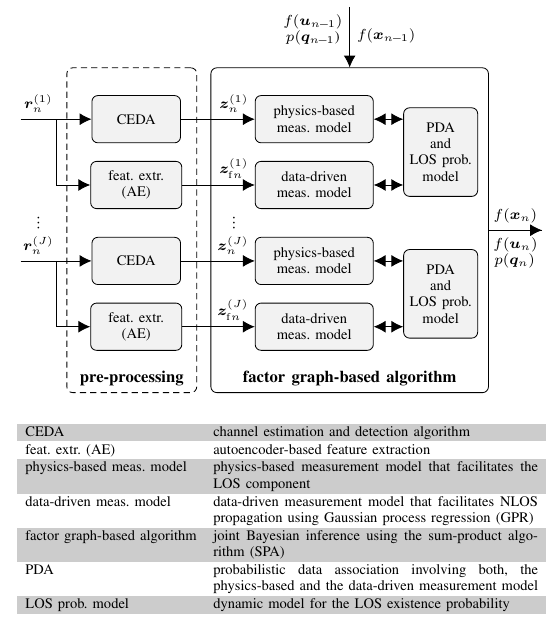}}
	\vspace{-4mm}
	\setlength{\abovecaptionskip}{0pt}
	\setlength{\belowcaptionskip}{0pt}
	\caption{Flowchart illustrating the main components of the proposed method. For each base station $j \rmv\rmv \in \rmv\rmv \{1,\s...\s,J\}$ we observe a baseband radio signal signal vector $\V{r}_n^{(j)}$ which is independently pre-processed into signal component measurements $\RV{z}_{n}^{(j)}$ (LOS component and \acp{mpc}) and feature measurements $\RV{z}_{\text{f}\s n}^{(j)}$. The factor graph-based algorithm uses these measurements and the distributions of agent state, amplitude state, and LOS probability state at the previous time step, given as ${f}(\V{x}_{n \minus 1})$, ${f}(\V{u}_{n \minus 1})$, ${p}(\V{q}_{n \minus 1})$, to infer the posterior distributions ${f}(\V{x}_n)$, ${f}(\V{u}_n)$, ${p}(\V{q}_n)$ at the current time step. For simplicity, we omit conditional dependencies of the distributions.}\label{fig:eye_catcher}
	\vspace{-6mm}
\end{figure}
Prominent examples of such approaches are multipath-based methods that take advantage of multipath by estimating \acp{mpc} and associating them to \acp{va} representing the location of the mirror image of a base station (anchor) on a reflecting surface. The locations of \acp{va} are assumed to be know a priori \cite{LeitingerGNSS2016} or estimated jointly with the position of the agent using \ac{mpslam} \cite{LeitingerTWC2019,GentnerTWC2016,KimTWC2020}. 
\ac{mpslam} provides high-accuracy position estimates even in \ac{olos} situations. However, it requires specular, resolved \acp{mpc} caused by environments consistent with the \ac{va} model, i.e, flat surfaces of sufficient extend \cite{PedersenJTAP2018,JiangOJAP2022,VenusTWC2023}. 
Other methods exploit cooperation among individual agents\cite{WymeerschProc2009,KulmerTWC2018,WangShenTWC2020}, or perform robust signal processing against multipath propagation and clutter measurements in general. 
The latter comprise heuristics \cite{DardariProcIEEE2009,ChianiProc2018}, machine learning-based approaches \cite{WymeerschIEEE2012,MaranoJSAC2010} as well as Bayesian methods\cite{MeyerFUSION2018,Yu2020,VenusRadar2021,VenusTWC2023}, and hybrids thereof \cite{Bartoletti2015, Mazuelas2018, ContiProcIEEE2019, WangShenTWC2020}. 
In particular, the methods introduced in \cite{WielandnerJAIF2023} and \cite{Yu2020,VenusRadar2021,VenusTWC2023} perform \ac{mpslam} or robust positioning considering delay dispersion and non-resolvable dense \ac{mpc} \cite{JiangOJAP2022}. 
In recent years, machine learning-based approaches %
have grown increasingly popular. Typically, they extract specific features from the radio channel, applying model-agnostic supervised regression methods on these features \cite{WymeerschIEEE2012,MaranoJSAC2010}.
While these approaches potentially provide high accuracy estimates at low computational demand (after training), 
they suffer from their dependence on a large representative measurement database and can fail in scenarios that are not sufficiently represented by the training data. This is why recent algorithms facilitate deep learning and autoencoder-based, unsupervised methods \cite{KinWel:ICLR2014, StahlkeSensors2021, WangWCOMLetter2021, LiShenMILCOM2021, KadShrAraDorWelAmiYerYoo:ICC2022, HuangTMC2022} or try to incorporate physics-based information in a systematic manner \cite{LiaMey:FUSION2022,LiaMey:SSP2021,Kram2022GPR,PinHessWymSven:TAES2023} to %
reduce the dependence on training data . 
Multipath-based localization \cite{VenusTWC2023, LeitingerTWC2019, LeitingerICC2019, GentnerTWC2016}, multiobject-tracking \cite{BarShalomTCS2009, MeyerProc2018, MeyerTSP2021}, and parametric channel tracking \cite{LiTWC2022} are applications that pose common challenges such as for example uncertainties beyond Gaussian noise, like missed detections and clutter, an uncertain origin of measurements, and unknown and time-varying numbers of objects to be localized and tracked. These challenges are well addressed by Bayesian inference leveraging graphical models to perform joint detection and estimation. In particular, the \ac{pda} algorithm \cite{BarShalomTCS2009,BarShalom1995} represents a low-complexity Bayesian method for robust localization and tracking with extension to multiple-sensors \ac{pda} \cite{JeoTugTAES2005} and amplitude information \cite{LerroACC1990}.

\subsection{Problem Statement and Contributions}
The problem studied in this paper can be summarized as follows.

\vspace{0.5mm}
	\noindent{\textit{Estimate the time-varying location of a mobile agent using \ac{los} propagation and multipath propagation of radio signals with emphasis on overcoming \ac{olos} situations.}}
\vspace{1.2mm}

We propose a neural-enhanced \ac{spa} for robust radio signal-based localization and tracking in multipath-prone envirnoments. %
The proposed algorithm performs joint probabilistic data association and sequential estimation of a mobile agent state together with all relevant model-parameters (amplitude state, LOS probability, data association variable) using message passing by means of the \ac{spa} on a factor graph \cite{KschischangTIT2001}.
The main components of the proposed algorithm are illustrated in the flowchart provided in Fig.~\ref{fig:eye_catcher}. We introduce a hybrid physics-based and data-driven model, which allows the proposed sequential algorithm to leverage the information contained in both, the LOS component and \acp{mpc} (NLOS components) of multiple base stations to robustly estimate the mobile agent's position.   
{Similar to other ``two-step approaches" \cite{LeitingerTWC2019, GentnerTWC2016, MeyerProc2018, MeyerTSP2021}, the proposed algorithm uses signal component measurements consisting of delays and corresponding amplitudes estimated out of the received baseband signal by a snapshot-based parametric \ac{ceda}. Additionally, our hybrid method uses feature measurements extracted out of the received baseband signal by an \ac{aednn} \cite{Kramer1991AE,Kram2022GPR}.}
Using the measurements provided by the \ac{ceda}, our physics-based model allows the algorithm to facilitate the position-related information contained in the LOS component with high accuracy and without the need of training data. The data-driven model, which is based on \ac{gpr}\cite{Rasmussen2006GP}, uses the feature measurements extracted by the \ac{aednn} to leverage the complex position-related information inherently contained in multipath components regardless of the source of multipath (e.g. flat walls and point scatters, but also curved and rough walls as well as irregular objects such as shelves or pillars). 
The introduced physics-based and data-driven models interact through a joint probabilistic data association model  \cite{MeyerProc2018,LeitingerTWC2019} and a dynamic LOS existence probability model \cite{SoldiTSP2019,VenusTWC2023}.
This allows the algorithm to robustly fuse the information contained in both, the LOS component and \acp{mpc} and, thus, to operate accurately and reliably within challenging environments, characterized by strong multipath propagation and \ac{olos} situations.
{\setlength\parindent{1em}
The key contributions of this paper are as follows:}
\begin{itemize}[leftmargin=1.5em]
	\item We introduce a novel Bayesian model for \ac{mpc}-aided sequential inference of the agent position consisting of a hybrid physics-based and data-driven measurement model. %
	\item We present an \ac{spa} based on the factor graph representation of the estimation problem, which efficiently infers all parameters of the introduced joint probabilistic model.
	\item We demonstrate that our algorithm robustly and accurately fuses the information contained the presented hybrid model. It outperforms state-of-the-art methods for \ac{nlos} mitigation \cite{WymeerschIEEE2012,Kram2022GPR,VenusTWC2023} and constantly attains the \ac{pcrlb} \cite{Tichavsky1998}.
\end{itemize}

\section{Overview} \label{sec:overview}
The problem considered is the sequential estimation of the agent state ${\RV{x}}_n$, while the agent is moving along an unknown trajectory. The current state of the agent is described by the state vector $\bm{x}_n = [\bm{p}_n ^\text{T}\; \bm{v}_n^\text{T} ]^\text{T}$, which is composed of the mobile agent's position $\bm{p}_n = [p_{\text{x}\s n}\; p_{\text{y}\s n}]^\text{T}$ and velocity $\bm{v}_n = [v_{\text{x}\s n}\; v_{\text{y}\s n}]^\text{T}$.
At each discrete time $n$, the mobile agent %
transmits a signal and each anchor (base station) $j \rmv\rmv \in \rmv\rmv \{1,\s...\s,J\}$ at anchor position $\bm{p}_{\text{A}}^{(j)} = [p_{\text{Ax}}^{(j)} \; p_{\text{Ay}}^{(j)}]^\text{T}$ acts as a receiver. For each anchor, we obtain the complex baseband signal vector $\bm{r}_n^{(j)} \in \mathbb{C}^{N_\text{s}}$ with ${N_\text{s}}$ being the number of signal samples.
Fig.~\ref{fig:eye_catcher} shows the main components of the proposed algorithm. In a pre-processing stage, we apply to $\bm{r}_n^{(j)}$ both, a \ac{ceda} and an \ac{aednn} to obtain signal component measurements $\V{z}_{n}$ and feature measurements $\V{z}_{\text{f}\s n}$, respectively. 
The agent state can be sequentially estimated in a Bayesian sense using all available measurements $\V{z}_{1:n}$, $\V{z}_{\text{f}\s 1:n} $ of all anchors up to time $n$ by using the \ac{mmse} estimator \cite{Kay1993} 
\begin{equation}\label{eq:mmse}
\hat{\bm{x}}^\text{MMSE}_{n} \,\triangleq \int \rmv \bm{x}_{n} \, f(\bm{x}_{n} | \V{z}_{1:n}, \V{z}_{\text{f}\s 1:n}  )\, \mathrm{d}\bm{x}_{n} \, .
\end{equation}
with  $\hat{\bm{x}}^{\text{MMSE}}_n  =  $ $ [\hat{\bm{p}}^{\text{MMSE T}}_n  $ $ \, \hat{\bm{v}}^{\text{MMSE T}}_n ]^\text{T}$ being the MMSE estimate. 

The proposed factor graph-based algorithm infers the marginal posterior distribution $ f(\bm{x}_{n} | \V{z}_{1:n}, \V{z}_{\text{f}\s 1:n}  )$ by executing the \ac{spa} on the factor graph that represents the hybrid probabilistic model introduced in this work. %
The structure of the following two sections, which present the introduced system model, aligns with the main components illustrated in Fig.~\ref{fig:eye_catcher}.

\section{Pre-Processing} \label{sec:pre_estimation}

\subsection{Channel Estimation and Detection} \label{sec:channel_estimation}
				
We independently apply, at each time $ n $ and for each anchor $j$, a parametric \ac{ceda} \cite{RichterPhD2005,BadiuTSP2017,LeitingerAsilomar2020,VenusTWC2023} to the complex baseband signal vector $\bm{r}_n^{(j)}$. The \ac{ceda}  decomposes $\bm{r}_n^{(j)}$ into individual components, yielding a number of $M_n^{(j)}$ measurements denoted by ${\V{z}^{(j)}_{m,n}}$ with $m \in \Set{M}_n^{(j)} \triangleq \{1,\,\dots\,,M_n^{(j)}\} $ that are collected by the vector $\vspace*{-0.2mm}\V{z}^{(j)}_{n} = [{\bm{z}^{(j)\text{T}}_{1,n}}  \rmv \cdots  {\V{z}^{(j)\text{T}}_{M_n^{(j)}\rmv\rmv\rmv\rmv,n}}]^\text{T}$. 
Each $\V{z}^{(j)}_{m,n} = [\zd~\zu]^\text{T}$ represents a signal component parameter estimate, containing a distance measurement ${z_\text{d}}_{m,n}^{(j)} \rmv\rmv\in\rmv\rmv [0, d_\text{max}]$ and a normalized amplitude measurement $\zu \in [\gamma, \infty)$, where $d_\text{max}$ is the maximum possible distance and $\gamma$ is the detection threshold of the \ac{ceda}, which is a constant to be chosen\footnote{Note that a low value for $\gamma$ results in an increased number of false alarms, but allows the detection of low amplitude \acp{mpc}. See \cite{LeitingerAsilomar2020} for determining $\gamma$ from the desired false alarm probability.}. Here, ``normalized amplitude'' refers to the square root of the \ac{snr} of the signal component. See \cite{LiTWC2022} for details on the signal model underlying a \ac{ceda}.  %
The stacked vector $\bm{z}_n = [\bm{z}^{(1)\, \text{T}}_{n} \rmv\rmv\rmv ... \, \bm{z}^{(J)\,\text{T}}_{n}]^\text{T}$ is used by the proposed algorithm as a noisy measurement. We also define the vector $\RV{M}_n = [\rv{M}_n^{(1)}\, ... \, \rv{M}_n^{(J)}]^\text{T}$. %

\subsection{AE-based Feature Extraction} \label{sec:feature_extraction}

At each time $n$ and for each anchor $j$ we independently apply a pre-trained \acf{aednn} \cite{Kramer1991AE,Kramer1992AE} where the complex baseband signal vector $\bm{r}_n^{(j)}$ acts as the decoded input\footnote{As suggested in \cite{Kram2022GPR}, we used the magnitudes of the complex baseband signal. However, our work could be extended to complex neural networks along the lines of \cite{Hirose2012Book,TrabelsiICLR2017}.}. We use the \ac{mse} between the training inputs and the predictions of the decoder network of the \ac{aednn} as a loss function for training. The latent (encoded) space of the \ac{aednn} consists of $F$ feature measurements denoted by $\zf$ collected by the vector  $\V{z}^{(j)}_{\text{f}\s n} = [{z^{(j)}_\text{f}}_{\hspace{-2.6mm} 1,n} \, ... \,  {z^{(j)}_\text{f}}_{\hspace{-2.6mm} F,n}]^\text{T} $, where $i$ is the feature index. Since we choose $F\ll {N_\text{s}}$, the \ac{aednn} can be said to compress the information contained in the received signal vector into $\V{z}^{(j)}_{\text{f}\s n}$. The encoder-decoder architecture of the \acp{aednn} enables unsupervised training using unlabeled samples of baseband signals collected in the set $\{ \bm{r}_{n'}^{(j)} \}_{n'=1}^{N'}$, where ${N'}$ is the number of training samples. The stacked vector $\bm{z}_{\text{f}\s n} = [\bm{z}^{(1)\, \text{T}}_{\text{f}\s n} \rmv\rmv\rmv ... \, \bm{z}^{(J)\,\text{T}}_{\text{f}\s n}]^\text{T}$ is used by the proposed algorithm as an additional noisy measurement.

\section{Factor Graph-based Algorithm} \label{sec:system_model}

\subsection{Random Variables and Assumptions} \label{sec:assumptions}
The current state of the agent $\bm{x}_n$ is the primary random variable to be inferred by the proposed factor graph-based algorithm. 
Additionally, we define the auxiliary state variables ${u}_n^{(j)}$, $\rv{q}_n^{(j)}$, and $\rv{a}_n^{(j)}$, which denote the normalized amplitude, LOS probability and association variable, respectively, and are modeled {separately} for all anchors. 
The presented measurement model consists of a physics-based measurement model, which uses the signal component measurements $\RV{z}_{n}^{(j)}$ obtained by the \ac{ceda}, and a data-driven measurement model, which uses the \ac{ae}-based feature measurements $\RV{z}_{\text{f}\s n}^{(j)}$. Our model is based on the assumptions that
\begin{itemize}
	\item[(I)] \ac{ceda}-based signal component measurements are uninformative with respect to \ac{nlos} components (\acp{mpc})\footnote{Assumption I is commonly used in \ac{pda}\cite{BarShalom1995} representing the least informative model for clutter measurements. Although it does not consider the precise statistics of measurements originating from \acp{mpc}, it does not affect the estimate as it does not impose curvature on the likelihood model.}.
	\item[(II)] \ac{ae}-based feature measurements are uninformative with respect to the \ac{los} component\footnote{Assumption II is not true in LOS situations, which leads to an overconfident estimate (reduced position uncertainty) of the agent posterior. However, this is counteracted by the LOS probability model that ``deactivates" the feature-based likelihood in LOS condition by causing the existence probability $\pe$ (see Sec.~\ref{sec:association_model}) to approach $1$.}.
	\item[(III)] \ac{ceda}-based measurements and feature-based measurements are conditionally independent for different values of $m$ \cite{MeyerProc2018,LeitingerTWC2019} and $i$ \cite{ZhaoVTC2016,Kram2022GPR} given the state variables.%
\end{itemize}

\subsection{Physics-based Measurement Model} \label{sec:los_model}
The \ac{los} \ac{lhf} of an individual distance measurement $\zdr$ is given by
\begin{equation}
	f_\text{L}(\zd | \bm{p}_n, u_n^{(j)}) \triangleq f_\text{N}(\zd;\, d_\text{LOS}^{(j)}(\bm{p}_n),\, \sigma_{\text{d}}(u^{(j)}_{n}))
\end{equation}
where $f_\text{N}(\cdot)$ denotes a Gaussian \ac{pdf} of the \ac{rv} $\zd$ with mean $d^{(j)}_{\text{LOS}}(\RV{p}_n)$ and variance $\sigma^2_\text{d}(\rv{u}^{(j)}_{n})$. The mean is physically related to the agent position via $d^{(j)}_{\text{LOS}}(\RV{p}_n) = \norm{\bm{p}_n - \bm{p}_\text{A}^{(j)}}{}$. The variance is determined by the Fisher information given by
$ \sigma_{\text{d}}^{2} (\rv{u}^{(j)}_{n}) =   c^2 / ( 8\,  \pi^2 \, \beta_\text{bw}^2 \, \rv{u}^{(j)\s 2}_{n} ) $ \cite{ShenTIT2010}, where $\beta_\text{bw}$ is the root mean squared bandwidth \cite{WitrisalSPM2016Copy,LeitingerJSAC2015} and $\rv{u}_n^{(j)}$ is the normalized amplitude \cite{LiTWC2022, VenusTWC2023}. 
The \ac{los} \ac{lhf} of the normalized amplitude measurement $\zur$ is modeled as \cite{LeitingerICC2019,LiTWC2022}
\begin{equation} \label{eq:los_amplitude}
	f_\text{L}(\zu| u^{(j)}_{n}) \triangleq f_\text{TRice}(\zu; \sigma_\mathrm{u} (u^{(j)}_{n}) ,u^{(j)}_{n}, \gamma)
\end{equation}
with $f_\text{TRice}(\cdot)$ being a truncated Rician PDF \cite{VenusTWC2023} with non-centrality parameter $u^{(j)}_{n}$ and  threshold value corresponding to $\gamma$ (see Sec.~\ref{sec:channel_estimation}). The scale parameter is again determined by the Fisher information and given as 
$\sigma_{\mathrm{u}}^{2} (\rv{u}^{(j)}_{n}) = 1/2 +\rv{u}^{(j)\s 2}_n /{(4\s N_{\text{s}})}$. %
See \cite{LiTWC2022} for a detailed derivation. %
Since we assume \ac{ceda}-based measurements to be uninformative with respect to \ac{nlos} propagation {(Assumption I)}, the \ac{nlos} \ac{lhf} of an individual distance measurement $\zdr$ is given as \cite{BarShalomTCS2009,BarShalom1995}
$%
f_\text{NL}(\zd) \triangleq f_\text{U}(\zd; 0,d_\text{max}) %
$ %
where $f_\text{U}(\cdot)$ denotes a uniform \ac{pdf} of the \ac{rv} \zd with the limits $0$ and $d_\text{max}$ corresponding to the distance measurement range of the \ac{ceda}. 
In correspondence to \eqref{eq:los_amplitude}, the NLOS \ac{lhf} of an individual normalized amplitude measurement $\zur$ is given as \cite{LeitingerICC2019,LiTWC2022}
$%
	f_\text{NL}(\zu) \triangleq  f_\text{TRayl}(\zu; \sqrt{1/2}, \gamma) %
$ %
where $f_\text{TRayl}(\cdot)$ is a truncated Rayleigh PDF with scale parameter of $\sqrt{1/2}$ and threshold value  corresponding to $\gamma$. 

\subsection{Data-driven Measurement Model} \label{sec:nlos_model}
The NLOS \ac{lhf} of individual feature measurements $\zf$ is modeled as
\begin{equation} \label{eq:nlos_feature_lhf}
	f_\text{NL}(\zf | \bm{p}_n) \triangleq f_\text{N}(\zf;\, \mu_\text{GP\s i}^{(j)}(\bm{p}_n),\, \sigma_{\text{GP}\s i}^{(j)}(\bm{p}_n) )
\end{equation}
with $\mu^{(j)}_{\text{GP}\s i}(\RV{p}_n)$ and $\sigma_{\text{GP}\s i}^{(j)\s 2}(\bm{p}_n)$ being the predicted mean and predicted variance, respectively, of a \acf{gpr} model\footnote{This choice leads to a computational complexity of the proposed method given as $\mathcal{O}(N''^{\s 2} J I F)$ \cite{Rasmussen2006GP} per time $n$, since the \ac{gpr}-based likelihood function in \eqref{eq:nlos_feature_lhf} must be evaluated for each feature, particle, and anchor, which leads to high runtimes for large sets of training data.} \cite{Rasmussen2006GP}. We train the \ac{gpr} model using \textit{labeled} data consisting of duples of feature measurements and according position labels collected in the set $\{(\bm{z}^{(j)}_{\text{f}\s n''}, \bm{p}_{n''}^{(j)})\}_{n''=1}^{N''}$, where ${N''}$ is the number of training samples.
Since we assume the \ac{ae}-based feature measurements $\zf$ to be non-informative with respect to \ac{los} propagation {(Assumption II)}, the \ac{los} \ac{lhf} of individual $\zf$ is given by
 $%
	f_\text{L}(\zf) \triangleq f_\text{U}(\zf; l^{(j)}_{\text{min}\s i}, l^{(j)}_{\text{max}\s i}),
$ %
where %
$l^{(j)}_{\text{min}\s i}$ and $l^{(j)}_{\text{max}\s i}$ are the lower and upper limits of the feature measurement range, respectively\footnote{For implementation, we normalize the latent variables using the ``pre-training'' dataset (see Sec.~\ref{sec:simulation_setup}) to obtain the feature-based measurements $\zf$, thus,  $l^{(j)}_{\text{min}\s i}= 0$ and  $l^{(j)}_{\text{max}\s i}= 1$ for all $i$ and $j$.}. %

\subsection{Probabilistic Data Association (PDA) Model} \label{sec:association_model}
At each time $n$ and for each anchor $j$, the \ac{ceda} measurements, i.e., the components of $\RV{z}_n^{(j)}$ are subject to data association uncertainty. Thus, it is not known which measurement $\RV{z}_{m,n}^{(j)}$ originated from the \ac{los}, or which one is due to a ``\ac{nlos} measurement'', i.e., a measurement originating from an \ac{mpc} or a \ac{fa}. Based on the concept of \acf{pda} \cite{BarShalom1995}, we define the association variable $\rv{a}^{(j)}_{n}$ as %
\begin{equation}\label{eq:association}
	{a}^{(j)}_{n} \rmv \rmv = \rmv 
	\begin{cases} 
		m \rmv\rmv \in \rmv\rmv \mathcal{M}_n^{(j)}  \rmv\rmv \rmv\rmv, &
		\text{$\bm{z}_{m,n}^{(j)}$ is the LOS measurement in $\bm{z}_n^{(j)}$} \\
		0 \, , & \text{there is no LOS measurement in $\bm{z}_n^{(j)}$}
	\end{cases}\, .
\end{equation}
Using no prior knowledge about the mean rate of \ac{nlos} measurements 
 (so called ``non-parametric model''\cite{BarShalom1995}), the joint \ac{pmf} of $\rv{a}^{(j)}_{n}$ and $\rv{M}_n^{(j)}$ can be shown to be proportional to the function \cite[Sec. 3.4.3]{BarShalom1995}
\begin{equation}  \label{eq:prior_association}
	h(a^{(j)}_{n}\rmv\rmv\rmv, M_n^{(j)} ;u_n^{(j)}\rmv\rmv\rmv,q_n^{(j)}) =
	\begin{cases} 
		\frac{\pe}{M_n^{(j)}} \, , \rmv\rmv&  a^{(j)}_{n} \in \mathcal{M}_n^{(j)} \\[2mm] 1\minus\pe\,, \rmv\rmv\rmv & a^{(j)}_{n} = 0
	\end{cases}
\end{equation}
where the \ac{los} \textit{existence} probability $\pe$ denotes the probability that there is a \ac{los} measurement for the current set of measurements. It is given as $\pe \triangleq \pdrv\, \rv{q}_n^{(j)} $ with $\rv{q}_n^{(j)}$ being the probability of the event that the \ac{los} is \textit{not} obstructed, which we refer to as \textit{\ac{los} probability}, and $\pdrv$ being the detection probability, i.e., the probability that at time $n$ and for anchor $j$ the agent generates a radio signal component whose amplitude is high enough so that it leads to an \ac{los} measurement. The LOS probability $\rv{q}_n^{(j)}$ is modeled as a discrete RV taking its values from the finite set $\mathcal{Q} = \{\lambda_1,\, ... \,, \lambda_Q\}$, where $\lambda_i \in (0,1]$. See \cite[Sec. IV-D]{VenusTWC2023} for details. 
We also define the joint vector $\RV{a}_n = [\rv{a}^{(1)}_{n}\, ... \, \rv{a}^{(J)}_{n}]^\text{T}$. 
\begin{figure}[t]
	\centering
	\setlength{\abovecaptionskip}{0pt}
	\setlength{\belowcaptionskip}{0pt}
	\tikzsetnextfilename{likelihoods}
	\hspace{-3mm}\includegraphics{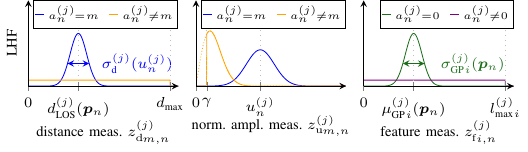}
	
	\caption{Graphical representation of the stochastic models constituting the overall \ac{lhf} for a single measurement.}
	\label{fig:single_measurment_like}
	\vspace{-5mm}
\end{figure}

Incorporating $\rv{a}^{(j)}_{n}$ into the model, we define the overall \ac{lhf} for individual distance measurements, given as 
\vspace{-1mm}
\begin{equation} \label{eq:single_delay_like}
	f(\zd | \RV{p}_n, \rv{u}_n^{(j)}, \rv{a}_n^{(j)})\rmv = \rmv
	\begin{cases} 
		\rmv f_\text{L}(\zd | \bm{p}_n,\rmv u_n^{(j)})  ,  \rmv\rmv &  a^{(j)}_{n} = m \\
		\rmv f_\text{NL}(\zd ) , \rmv\rmv  & a^{(j)}_{n} \neq m
	\end{cases}
\end{equation}
and the overall \ac{lhf} for individual normalized amplitude measurements, given as 
\vspace*{-1mm}
\begin{equation}
	f(\zu |u^{(j)}_{n}, a^{(j)}_{n} ) =\begin{cases} 
		\rmv f_\text{L}(\zu| u^{(j)}_{n})  , \rmv\rmv\rmv &  a^{(j)}_{n} = m \\
		\rmv	f_\text{NL}(\zu ) , \rmv\rmv\rmv  & a^{(j)}_{n} \neq m  
	\end{cases} \, . \label{eq:single_ampl_like} 
\end{equation}
We seek to utilize the information contained in the data-driven model (i.e., \ac{nlos} information) only in situations where the \ac{los} is not available. Therefore, we define the overall \ac{lhf} for individual feature measurements as 
\begin{equation}
	f(\zf | \RV{p}_n, a^{(j)}_{n} ) =\begin{cases} 
		\rmv 	f_\text{NL}(\zf | \bm{p}_n)  , \rmv\rmv\rmv &  a^{(j)}_{n} = 0\\
		\rmv	f_\text{L}(\zf ) , \rmv\rmv\rmv  & a^{(j)}_{n} \neq 0 
	\end{cases}  \, . \label{eq:single_feature_like} 
\end{equation}
The shapes of \eqref{eq:single_delay_like}\,, \eqref{eq:single_ampl_like}\,, and \eqref{eq:single_feature_like} are depicted in Fig.~\ref{fig:single_measurment_like}. 
By assuming conditional independence of \ac{ceda}-based measurements and feature-based measurements (Assumption I and II) and conditional independence of measurements
for different values of $m$ and $i$ (Assumption III), the joint \ac{lhf} for all measurements per anchor $j$ and time $n$ is given as
\begin{align} \label{eq:overall_lhf}
	&f(\bm{z}_n^{(j)}\rmv\rmv\rmv, \V{z}^{(j)}_{\text{f}\s n} | \RV{p}_n, \rv{u}_n^{(j)}\rmv\rmv\rmv, \rv{a}_n^{(j)}) \rmv \nonumber\\ 
	&\hspace{9mm} =  \big(\rmv\rmv\rmv \prod_{m=1}^{\,\,M_{n}^{(j)}} \rmv\rmv\rmv 	f(\zd | \RV{p}_n, \rv{u}_n^{(j)}\rmv\rmv\rmv, \rv{a}_n^{(j)}) \, f(\zu |u^{(j)}_{n}\rmv\rmv\rmv, a^{(j)}_{n} )	\big) \nonumber\\ 
	& \hspace{12mm} \times \big( \prod_{i=1}^{\,\,F} 	f(\zf | \RV{p}_n, a^{(j)}_{n}) \big) \, .
\end{align}

\subsection{State Transition Model} \label{sec:state_transition}
We model the evolution over time $n$ of $\RV{x}_n$ and $u_n^{(j)}$ and $\rv{q}_n^{(j)}$ as first-order Markov processes which are distributed independently, i.e.,
$ f({\bm{x}}_n, \bm{u}_n,  \bm{q}_{n} |{\bm{x}}_{n\minus 1}, \bm{u}_{n\minus 1},  \bm{q}_{n\minus 1})%
= f({\bm{x}}_n |{\bm{x}}_{n\minus 1}) $ $ \prod_{j=1}^{J} $  $f({u}_n^{(j)}| {u}_{n\minus 1}^{(j)}) $ $p(q_n^{(j)}| q_{n\minus 1}^{(j)}) $
 with 
  the joint vectors
 $\RV{u}_n = [\rv{u}^{(1)\s \text{T}}_{n} \vspace{0.5mm}\, ... \, \rv{u}^{(J) \s \text{T}}_{n}]^\text{T}$, and
 $\RV{q}_n = [\rv{q}^{(1)}_{n}\, ... \, \rv{q}^{(J)}_{n}]^\text{T}$.  %

\subsection{Joint Posterior and Factor Graph}  \label{sec:factor_graph}
\begin{figure}[t]
	\centering
	%
	%
	%
	%
		\scalebox{0.92}{\includegraphics{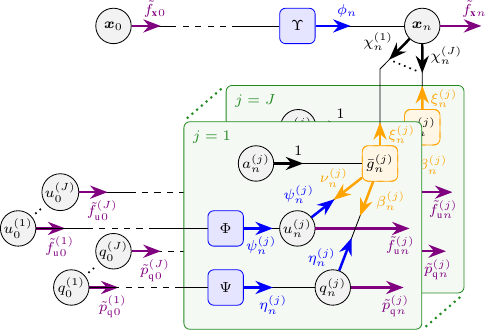}}%
	\vspace*{-2.5mm}
	\caption{Factor graph representing the factorization of the joint posterior \ac{pdf} in \eqref{eq:factorization1} and the respective messages according to the SPA. The following short notations are used for the marginal posterior messages: %
		$ \breve{f}_{{\text{\textbf{x}}}\s n}        \rmv\triangleq\rmv  \breve{f}_{{\text{\textbf{x}}}}(\bm{x}_{n})    $, 
		$ \breve{f}_{\text{\textbf{y}}\s n}^{(j)}        \rmv\triangleq\rmv  \breve{f}_{\text{\textbf{y}}}(u_{n}^{(j)})          $, 
		$ \breve{p}_{\text{q}\s n}^{(j)}                 \rmv\triangleq\rmv  \breve{p}_{\text{q}}(q_{n}^{(j)})                        $. See \cite[Sec. VI]{VenusTWC2023} for details regarding the messages
		$ \eta_n         $,
		$ \phi_n^{(j)}   $,
		$ \psi_n^{(j)}   $, 
		$ \xi^{(j)}_n    $,
		$ \chi^{(j)}_n   $,
		$ \nu_n^{(j)}    $,
		$ \beta_n^{(j)}  $,  and
		$ \chi^{(j)}_n   $		
		.
	}\label{fig:factor_graph}
	\vspace{-3mm}
\end{figure}
Let 
$\bm{z}_{1:n} \rmv \rmv\rmv\rmv = \rmv\rmv\rmv  [\bm{z}^\text{T}_{1}\, ... \, \bm{z}^\text{T}_{n}]^\text{T} \rmv\rmv\rmv \rmv\rmv\rmv  $ , 
$\V{z}_{\text{f}\s 1:n} \rmv \rmv\rmv\rmv = \rmv\rmv\rmv  [\bm{z}^\text{T}_{\text{f}\s 1}\, ... \, \bm{z}^\text{T}_{\text{f} n}]^\text{T} \rmv\rmv\rmv \rmv\rmv\rmv  $ , 
$\bm{x}_{1:n} \rmv\rmv\rmv\rmv =  \rmv\rmv\rmv  [\bm{x}^\text{T}_{1}\, ... \, \bm{x}^\text{T}_{n}]^\text{T} \rmv\rmv\rmv $, 
$\bm{a}_{1:n} \rmv\rmv \rmv\rmv  =  \rmv\rmv\rmv   [\bm{a}^\text{T}_{1}  \, ... \, \bm{a}^\text{T}_{n}]^\text{T} \rmv\rmv\rmv $, 
$\bm{u}_{1:n}  \rmv\rmv \rmv\rmv =  \rmv\rmv\rmv [\bm{u}^\text{T}_{1}\, ... \, \bm{u}^\text{T}_{n}]^\text{T} \rmv\rmv\rmv $, 
$\bm{q}_{1:n}  \rmv\rmv \rmv\rmv =  \rmv\rmv\rmv [\bm{q}^\text{T}_{1}\, ... \, \bm{q}^\text{T}_{n}]^\text{T} \rmv\rmv\rmv $, 
$\bm{M}_{1:n}  \rmv\rmv \rmv\rmv =  \rmv\rmv\rmv [\bm{M}^\text{T}_{1}\, ... \, \bm{M}^\text{T}_{n}]^\text{T} \rmv\rmv\rmv $.
By applying Bayes' rule, the joint posterior \ac{pdf} of all state variables up to time $n$ and all $J$ anchors is given (up to irrelevant constant terms) as %
\vspace{-1mm}
\begin{align} \label{eq:factorization1}
&  f(\bm{x}_{1:n}, \rmv\bm{a}_{1:n}, \rmv\bm{u}_{1:n}, \rmv\bm{q}_{1:n}, \rmv\bm{M}_{1:n} | \bm{z}_{1:n}, \V{z}_{\text{f}\s 1:n}  ) 
\nonumber\\
&\propto f( \bm{x}_0) \rmv\rmv \prod^{J}_{j= 1} p( q_0^{(j)}) \,f( u_0^{(j)})  \rmv
  \rmv\rmv\prod^{n}_{n'= 1}  \rmv\rmv\rmv \Upsilon( \bm{x}_{n'} | \bm{x}_{n'\minus1})\, \Phi( u_{n'}^{(j)} | u_{n'\minus1}^{(j)}) \nonumber \\& ~~~~~\times \Psi( q_{n'}^{(j)} | q_{n'\minus1}^{(j)}) \,  \bar{g}( \bm{z}_{n'}^{(j)}\rmv\rmv\rmv, \V{z}^{(j)}_{\text{f}\s n'}; \bm{p}_{n'}, u^{(j)}_{n'} ,  a^{(j)}_{n'},  q^{(j)}_{n'})\,,
\end{align}
where we introduced the state-transition functions   $\Upsilon({\bm{x}}_n|{\bm{x}}_{n-1}) \triangleq f({\bm{x}}_n|{\bm{x}}_{n-1})$, $\Phi(u_n^{(j)}|u_{n-1}^{(j)}) \triangleq f(u_n^{(j)}|u_{n-1}^{(j)})$, and $\Psi(q_n^{(j)}| q_{n \minus 1}^{(j)}) \triangleq p(q_n^{(j)}| q_{n \minus 1}^{(j)})$. 
We also introduced the pseudo \acl{lhf} 
$\bar{g}(\bm{z}_n^{(j)}\rmv\rmv\rmv\rmv\rmv, \V{z}^{(j)}_{\text{f}\s n} ; \bm{p}_{n},\rmv u^{(j)}_{n} \rmv\rmv\rmv\rmv,  a^{(j)}_{n}\rmv\rmv\rmv\rmv, \rmv q_n^{(j)})
\rmv\rmv \triangleq h(a_{n}^{(j)} \rmv ; u_n^{(j)},q_n^{(j)})\,
g( \bm{z}_n^{(j)}\rmv\rmv\rmv, \V{z}^{(j)}_{\text{f}\s n}\rmv ; \bm{p}_n, \rmv u^{(j)}_{n}\rmv\rmv\rmv\rmv\rmv , \rmv a^{(j)}_{n})$, %
where we define \vspace{-1mm}
\begin{align}\label{eq:likelihood}
	&g(  \bm{z}_n^{(j)}\rmv\rmv\rmv, \V{z}^{(j)}_{\text{f}\s n}\rmv ; \bm{p}_n, u^{(j)}_{n} \rmv\rmv\rmv ,  a^{(j)}_{n}) \propto f(\bm{z}_n^{(j)}\rmv\rmv\rmv, \V{z}^{(j)}_{\text{f}\s n} | \RV{p}_n, \rv{u}_n^{(j)}\rmv\rmv\rmv, \rv{a}_n^{(j)})  \nonumber  \\ & \hspace{6mm}
	=  \rmv\rmv
	\begin{cases} 
		\prod_{i=1}^{\,\,F} f_\text{NL}(\zf | \bm{p}_n)  /  f_\text{L}(\zf) , &  a^{(j)}_{n} = 0 \\
		\Lambda( \bm{z}_{a^{(j)}_{n}\rmv\rmv\rmv, n}^{(j)} | \bm{p}_n, u^{(j)}_{n} )  , & a^{(j)}_{n} \in \mathcal{M}_n^{(j)}
	\end{cases} \\[-7mm] \nonumber
\end{align} 
by neglecting the constant terms in \eqref{eq:overall_lhf}$\s$, where
\vspace{-1mm}
\begin{equation} \label{eq:likelihood_ratio} \vspace{-1mm}
	\rmv \Lambda( \bm{z}_{m,n}^{(j)} | \bm{p}_n, u^{(j)}_{n} ) = \frac{	f_\text{L}(\zd | \bm{p}_n,	\rmv  u_n^{(j)})  \, 	f_\text{L}(\zu| u^{(j)}_{n}) }{f_\text{NL}(\zd)\, f_\text{NL}(\zu) }
\end{equation}
is the likelihood ratio of signal component measurements\footnote{For  $\rv{a}^{(j)}_{n} = 0$, \eqref{eq:likelihood} is determined by the likelihood ratio of the feature measurements and, thus, information is gained from the data-driven model. When $\rv{a}^{(j)}_{n} = m \in \mathcal{M}_n^{(j)}$ (not zero), \eqref{eq:likelihood} is determined by the likelihood ratio of signal component measurements and, thus, information is gained from the physics-based model w.r.t. the $m$-th signal component measurement.}. 
Note that $\bm{M}_{1:n}$ is fixed and thus constant, as it is defined implicitly by the measurements $\bm{z}_{1:n}$, thus, $h(a_{n}^{(j)} \rmv ; u_n^{(j)},q_n^{(j)}) \equiv h(a_{n}^{(j)}, M_n^{(j)} \rmv ; u_n^{(j)},q_n^{(j)})$. 
The joint posterior \ac{pdf} in \eqref{eq:factorization1} is represented by the factor graph shown in Fig.~\ref{fig:factor_graph}.

\subsection{Algorithm}  \label{sec:algorithm}
\begin{figure}[t]
	
	\centering
	\setlength{\abovecaptionskip}{0pt}
	\setlength{\belowcaptionskip}{0pt}
	
	\setlength{\figurewidth}{0.28\textwidth}
	\setlength{\figureheight}{0.28\textwidth}
	\scalebox{1}{\includegraphics{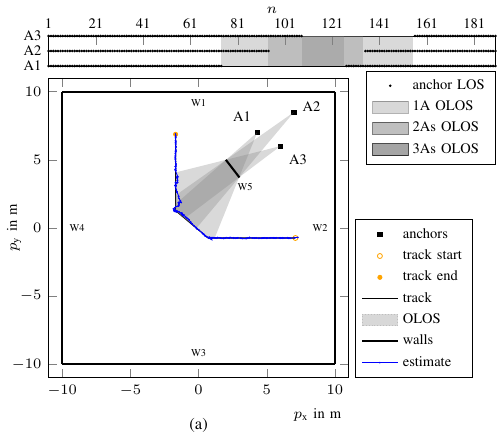}}
	\vspace{1mm}
	\hspace{-1mm}\includegraphics{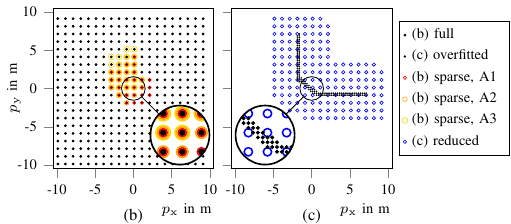}
	\caption{Graphical representation of the investigated synthetic experiment: Fig. (a) shows the simulated trajectory, anchor positions, walls, and OLOS intervals. Figs. (b) and (c) show the positions of all simulated training datasets. %
	}\label{fig:track_geometric}
	\vspace{-6mm}
\end{figure}
The problem considered is the sequential estimation of the agent state  $\hat{\bm{x}}^{\text{MMSE}}_n $ through the MMSE estimate given by \eqref{eq:mmse}. Furthermore, we also calculate MMSE estimates of normalized amplitude
$%
\hat{u}^{(j)\s\text{MMSE}}_{n} \rmv\rmv\rmv\rmv \triangleq \rmv\rmv\rmv\rmv  \int \rmv u_n^{(j)} \rmv\rmv\rmv f(u_n^{(j)} |    \V{z}_{1:n}, \V{z}_{\text{f}\s 1:n} )  \mathrm{d} u_n^{(j)}$ and LOS probability $ %
\hat{q}^{(j)\s\text{MMSE}}_{n} \rmv\rmv\rmv  \triangleq  \rmv\rmv\rmv \sum_{\omega_i \in \mathcal{Q}}   \rmv \omega_i \, p(q_n^{(j)} \rmv\rmv\rmv\rmv   = \rmv\rmv  \omega_i | \V{z}_{1:n}, \V{z}_{\text{f}\s 1:n} )
\, %
$. %
In order to obtain these quantities, the respective marginal posterior \acp{pdf} need to be calculated from the joint posterior \ac{pdf} in \eqref{eq:factorization1}. In general this is computationally infeasible\cite{MeyerProc2018}. Therefore, we use message passing by means of the \ac{spa} rules \cite{KschischangTIT2001} on the factor graph in Fig.~\ref{fig:factor_graph} that represents a factorization of the joint posterior \ac{pdf}. 
Since the integrals involved in the calculations of the \ac{spa} messages cannot be obtained analytically, we use a computationally efficient, sequential particle-based implementation that provides approximate results. See \cite[Sec. VI]{VenusTWC2023} for details 
concerning \ac{spa}, particle-based implementation and the determination of the initial distributions, i.e., $ f(\bm{x}_0), f(u_0^{(j)}) $, $p(q_0^{(j)})$
\footnote{The factorization structure given in \cite[Sec. VI]{VenusTWC2023} is identical to the problem at hand, when replacing $\bar{\bm{x}}_n$ with ${\bm{x}}_n$ and $\bm{y}_n^{(j)}$ with $u_n^{(j)}$. We introduce the additional approximation that the LOS existence probability $\pe $ is only affected by the physics-based model, i.e., when evaluating the messages $\nu(u_n^{(j)}) $, and $\beta(q_n^{(j)})$, we set $f(\zf | \RV{p}_n, a^{(j)}_{n}) \triangleq 1$ for all $i$.}.
Note that   
$  \breve{f}_{{\text{\textbf{x}}}}(\bm{x}_{n})    \propto   f(\bm{x}_{n} | \V{z}_{1:n}, \V{z}_{\text{f}\s 1:n}  )  $, 
$  \breve{f}_{\text{\textbf{y}}}(u_{n}^{(j)})      \propto     f(u_n^{(j)} |    \V{z}_{1:n}, \V{z}_{\text{f}\s 1:n} )       $, 
$  \breve{p}_{\text{q}}(q_{n}^{(j)})      \propto     p(q_n^{(j)}| \V{z}_{1:n}, \V{z}_{\text{f}\s 1:n}  )                  $ in Fig.~\ref{fig:factor_graph} denote the messages corresponding to the marginal posterior distributions. %
\section{Computational Results}\label{sec:results}

\subsection{Simulation Setup and Scenario} \label{sec:simulation_setup}
We evaluate the proposed algorithm using synthetic radio measurements, generated according to the scenario presented in Fig.~\ref{fig:track_geometric}a, where the agent moves along a trajectory with two distinct direction changes. It is observed at $190$ discrete time steps %
at a constant observation rate of $\Delta T = 100\,\mathrm{ms}$. %
The ground truth \ac{va} positions and corresponding \ac{mpc} distances are calculated based on the floor plan of Fig.~\ref{fig:track_geometric}a (W1 to W4) using the image source model \cite{PedersenJTAP2018,Meissner2015Diss}. 
The normalized amplitudes of the LOS component as well as the \acp{mpc} are assumed to follow free-space path loss according to their individual propagation paths, and are set to 38 dB at a distance of $1$ m. The normalized amplitudes of \acp{mpc} are additionally attenuated by 3 dB per reflection. 
The anchors are obstructed by an obstacle (W5), which leads to partial and full \ac{olos} situations in the center of the track.
We choose the transmitted signal to be of root-raised-cosine shape with a roll-off factor of $0.6$ and a $3$-dB bandwidth of $500\,\mathrm{MHz}$. The received baseband signal is critically sampled, i.e., $T_\text{s} = 1.25\,\mathrm{ns}$, with a total number of $N_\text{s} =81$ samples, amounting to a maximum distance $d_\text{max} = 30\,\mathrm{m}$. 
We use the \ac{ceda} from the supplementary material of \cite{VenusTWC2023} with detection threshold of $\gamma = 2$ (corresponding to $6\,\mathrm{dB}$). The \ac{aednn} is set up as suggested in \cite{Kram2022GPR}: We use feed-forward networks with three convolutional layers for both, encoder and decoder. The encoder is set up as $27\times{}17-\text{ELU}, 27\times{}13-\text{ReLU}, 16\times{}5-\text{ELU}$, which denotes the number of convolutional kernels times filter size and the respective activations, and applies max pooling of size 2 after all activation functions. It uses the magnitudes of the baseband signal vector $|\bm{r}_n^{(j)}|$ as an input and has a latent space of $4$ variables. The decoder network mirrors the encoder network. 
For implementation we used Python along with TensorFlow/Keras and optimized using Adam with learning rate of $2\cdot 10^{-3}$, using the \ac{mse} of measured and predicted values of $\bm{r}_n^{(j)}$ as loss function. To implement the feature measurement model of \eqref{eq:nlos_feature_lhf} we utilized MATLAB's \ac{gpr} toolbox, where we employed the ``Matern52" kernel function \cite{Kram2022GPR}. 
The state transition \ac{pdf} $f(\bm{x}_n|\bm{x}_{n-1})$ of the agent state $\RV{x}_n$ is described by a linear, constant-velocity and stochastic-acceleration model\cite[p.~273]{BarShalom2002EstimationTracking}, given as $\RV{x}_n = \bm{A}\, \RV{x}_{n\minus 1} + \bm{B}\, \RV{w}_{n}$ 
with 
the acceleration process $\RV{w}_n$ being i.i.d. across $n$, zero mean, and Gaussian with covariance matrix ${\sigma_{\text{a}}^2}\, \bm{I}_2$, %
the acceleration standard deviation ${\sigma_{\text{a}}}$, and $\bm{A} \in \mathbb{R}^{\text{4x4}}$ and $\bm{B} \in \mathbb{R}^{\text{4x2}}$ being defined according to \cite[p.~273]{BarShalom2002EstimationTracking}.
The state transition of the normalized amplitude $\rv{u}_{n}$, i.e., the state transition \ac{pdf} $f({u}_n|{u}_{n-1})$, is chosen as ${u}_n^{(j)} = {u}_{n\minus 1}^{(j)} + {\epsilon}_{\text{u}\s n}^{(j)}$, where the noise $ {\epsilon}_{\text{u}\s n}^{(j)}$ is i.i.d. across $n$, zero mean, Gaussian, with variance $\sigma^2_{\mathrm{u}}$.
The state transition variances are set as $\sigma_a=2~\mathrm{m/s^2}$ and  $\sigma_\text{u} = 0.05\,\hat{u}_{n\minus 1}^{(j)\s \text{MMSE}}$. The set of possible \ac{los} probabilities $\mathcal{Q}$ and the elements of the state transition \ac{pmf} $p(q_n^{(j)} | q_{n \minus 1}^{(j)} )$ are set in accordance to \cite[Sec. VII-A]{VenusTWC2023}.
The number of particles to represent the ``stacked state"\cite[Sec. VI]{MeyerProc2018,LeitingerTWC2019,VenusTWC2023} consisting of all random variables dependent over time $n$ was set to $I=5000$.
\begin{figure}[t!]
	
	\centering
	\setlength{\abovecaptionskip}{0pt}
	\setlength{\belowcaptionskip}{0pt}
	\tikzsetnextfilename{single_realization}
	\vspace{-1mm}
	\includegraphics{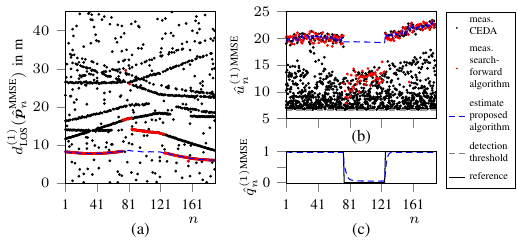}
	\vspace{-1mm}
	\caption{A single measurement realization and the respective \ac{mmse} estimates of the proposed algorithm in (a) distance domain $ d^{(1)}_{\text{LOS}} (\hat{\bm{p}}^{\text{MMSE}}_n)$, (b) amplitude domain $\hat{u}_n^{(1)\s \text{MMSE} \; 2}$, and (c) LOS probability domain $\hat{q}^{(1)\s\text{MMSE}}_{n}$. }\label{fig:single_realization}
	\vspace{-3mm}
\end{figure}
\begin{figure*}[t]
	
	\centering
	\setlength{\abovecaptionskip}{0pt}
	\setlength{\belowcaptionskip}{0pt}
	\tikzsetnextfilename{ml_features}
	\includegraphics{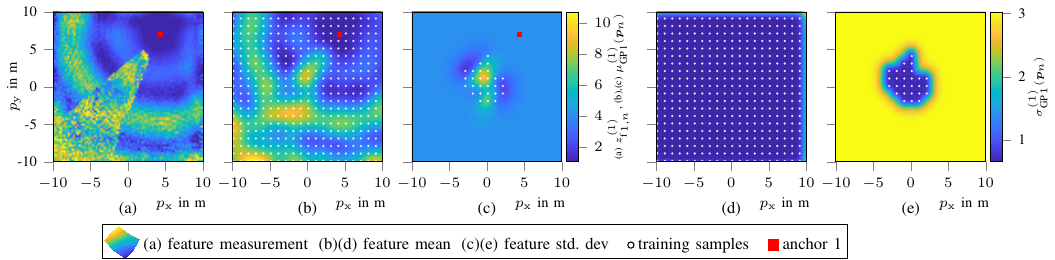}
	\caption{Exemplary illustration of the latent space representation learned by the \ac{aednn} in Fig. (a) along with the mapping learned by \ac{gpr} for feature $i=1$ of anchor $j=1$, where Figs. (b) and (c) represent the \ac{gpr} mean and Figs. (d) and (e) represent the \ac{gpr} standard deviation for the two datasets shown in Fig.~\ref{fig:track_geometric}b that are used for training of the proposed algorithm. The ``full" dataset is used for Figs. (b) and (d) and the ``sparse" dataset for Figs. (c) and (e).}\label{fig:ml_features}
	\vspace*{-3mm}
\end{figure*}
\vspace{-1mm}
\subsection{Reference Methods}
We compare the performance of the proposed method to a particle-based variant of the multi-sensor \ac{pdaai} \cite{JeoTugTAES2005}, the \ac{nlos} cluster based algorithm as presented in \cite{VenusTWC2023}, and the machine learning based methods from \cite{Kram2022GPR} and \cite{WymeerschIEEE2012}. 
Since the methods from \cite{Kram2022GPR} and \cite{WymeerschIEEE2012} do not perform data association, we estimate the LOS component distance using a search-forward approach \cite{DardariProcIEEE2009}. On the interpolated Bartlett spectrum \cite{KrimVibergSPM1996}, we search for the first maximum that exceeds a threshold, which we chose as six times the noise variance of the received baseband signal. 
The search-forward approach enables correctly identifying the LOS component (i.e., the first visible signal component), even when there are \acp{mpc} with amplitudes higher than that of the \ac{los} component. For the method from \cite{Kram2022GPR} we set up the \ac{aednn} and the \ac{gpr} with identical configurations as the proposed algorithm (corresponding to the suggestions in \cite{Kram2022GPR}). Initial distribution as well as state transition model of the agent state were also set in accordance with the proposed algorithm. To ensure fair comparison, we used Fisher information based variances for the delay likelihood model instead of heuristically set values. Note that this method performs ``anomaly detection", i.e, a data-driven identification of \ac{olos} situations, using a beta variational \ac{aednn}\cite{KinWel:ICLR2014} that is implemented using the configuration suggested in \cite{Kram2022GPR,StahlkeSensors2021}\footnote{The authors suggest to use feed-forward networks with three dense layers for both, encoder and decoder. The encoder uses the stacked real and imaginary parts of the baseband signal vector as an input. It consists of 100, 80, and 60 neurons, respectively, all with ReLU activation functions, and it has two latent variables. The decoder mirrors the encoder. The regularization hyper parameter is set to $\beta=10^{-3}$ and the \ac{mse} is used as a data reconstruction loss. As suggested, we used the ``time index signal strength indicator" for predicting the anomaly score and compared to the optimum detection threshold being set to the intersection point of the histograms of the agent trajectory data (which is not available in reality).}.
For the method from \cite{WymeerschIEEE2012}, we provide results using the setup referred to as ``GP", which learns a bias correction term using \ac{gpr} based on the six parametric features suggested by the authors.\footnote{Note that the approach based on support vector machines (termed ``SVM" in \cite{WymeerschIEEE2012}) did not yield stable results for the investigated experiment. Using logarithmic features (``log-GP") also did not improve the results, while this variant is prohibitive when negative bias values occur.} After error correction of the distance measurements according to \cite{WymeerschIEEE2012}, for fair comparison, we infer the agent position using a particle filter with identical configuration (initial distributions, state transition model) as the proposed method instead of using the suggested maximum likelihood positioning. Also, in accordance with the proposed method, the likelihood variances are determined from the Fisher information. 
\vspace{-2mm}
\subsection{Training procedure} \label{sec:training}
\vspace{-1mm}
Training of the proposed algorithm involves a two step procedure. First, the \ac{aednn} learns a low dimensional latent representation of the received signal (``pre-training"). We used $N'=6400$ unlabeled samples that cover the entire floorplan constituting a two-dimensional grid from $-10$~m to $10$~m in $p_\text{x}$ and $p_\text{y}$ directions with $0.25$~m spacing. Second, \ac{gpr} is used to learn the mapping in \eqref{eq:nlos_feature_lhf}. We provide results for two sets of training data as depicted in Fig.~\ref{fig:track_geometric}b: A dataset covering the entire floorplan with $1$~m grid spacing (``full") and a small dataset covering only the OLOS regions of the respective anchors (``sparse"). 

While the multi-sensor \ac{pdaai} and the algorithm from \cite{VenusTWC2023} require no training, the method from \cite{Kram2022GPR} was trained using the same training data as the proposed algorithm. However, this method additionally requires training of the variational \ac{aednn} used for ``anomaly detection" with LOS only data (i.e., no OLOS situations). We used $6400$ samples generated at the same positions as for ``pre-training" while deactivating the obstructing wall (W5). 
Training of the method from \cite{WymeerschIEEE2012} requires only one set of training data labeled with their respective positions. Here, we instead provide results using the two datasets depicted in Fig.~\ref{fig:track_geometric}c. The ``reduced" dataset consists of those positions of the ``full" dataset, where the overall received signal power remains within a moderate range. We found a low received signal power to be detrimental for this method leading to strong fluctuations of the distance error for adjacent positions. Additionally, we used an ``overfitted" dataset, which contains only data located around the true  agent trajectory. 

\subsection{Numerical Results and Performance Analysis}
Fig.~\ref{fig:single_realization} shows a single measurement realization and respective MMSE estimates of the proposed method. We show measurements obtained using both, the \ac{ceda} and the search-forward approach used for the comparison methods.
Fig.~\ref{fig:ml_features} gives an exemplary illustration of the latent space representation learned by the \ac{aednn} as well as the corresponding mapping of mean and standard deviation learned by \ac{gpr} as a function of the agent position $\bm{p}_n$ for the ``full" and the ``sparse" dataset (see Sec.~\ref{sec:training}). It can be observed that the \ac{gpr} mean values in Fig.~\ref{fig:ml_features}b and Fig.~\ref{fig:ml_features}c align well with the abstract feature space. The \ac{gpr} standard deviations in Fig.~\ref{fig:ml_features}d and Fig.~\ref{fig:ml_features}e remain consistently low in the learned regions and increase significantly in areas where no training data is available. 
Fig.~\ref{fig:results} shows the results of the performed numerical simulation. 
The results are shown in terms of both, the \ac{rmse} of the estimated agent position over time $n$ given as  $e_{n}^{\text{RMSE}}~=~\sqrt{\E{\norm{\hat{\bm{p}}^{\text{MMSE}}_n -\bm{p}_n}{2}}}$ and the cumulative frequency of the magnitude error of the estimated agent position, and are evaluated using a numerical simulation with 500 realizations. 
 \newcommand{\myopa}{1}
 \newcommand{\myopb}{1}
 \newcommand{\myopc}{1}
 \newcommand{\myopd}{1}
 \newcommand{\myope}{1}
 \newcommand{\myopf}{1}
 \newcommand{\myopg}{1}
 \newcommand{\shadesofgraysynthetic}{Different shades of gray represent different numbers of anchors in OLOS according to Fig.~\ref{fig:track_geometric}.}
 \begin{figure}[t]
	\centering
	\setlength{\belowcaptionskip}{0pt}	
	\tikzsetnextfilename{mse_all}
    \includegraphics{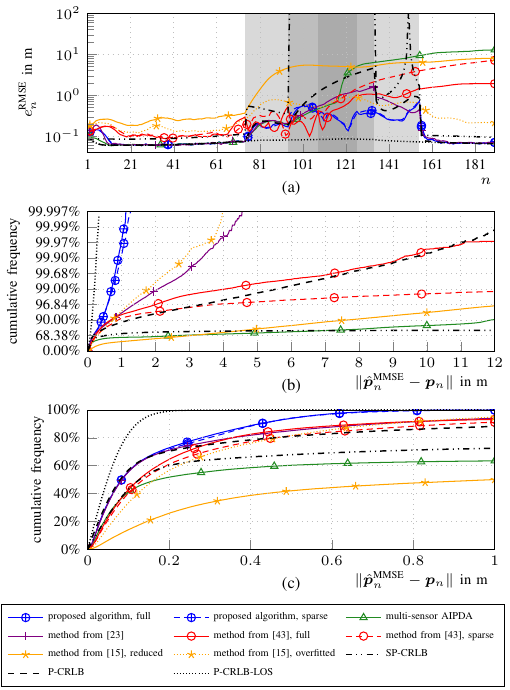}
	\caption{Performance in terms of the \ac{rmse} of the estimated agent position over time $n$ (a) and as the cumulative frequency of the magnitude error of the estimated agent position in inverse logarithmic scale (b) and linear scale (c). \shadesofgraysynthetic
	}\label{fig:results}
	\vspace{-6mm}
\end{figure}

As a performance benchmark, we provide the \ac{crlb} for a single position measurement without tracking (``SP-CRLB") as well as the \ac{pcrlb}, considering the dynamic model of the agent state \cite{Tichavsky1998, VenusTWC2023}.  The ``P-CRLB-LOS" is the P-CRLB assuming the LOS component to all anchors is always available. 

The \ac{rmse} of the proposed algorithm attains or even outperforms the \ac{pcrlb} during full \ac{olos} due to the additional information provided by the geometric imprint of the \ac{nlos} components and shows consistent performance for both, ``full" and ``sparse" traning data. While the cluster based approach from \cite{VenusTWC2023} also reaches the \ac{pcrlb} in LOS condition and manages to maintain the track in every single realization, it shows reduced performance during the \ac{olos} situation. 
The method from \cite{Kram2022GPR} cannot attain the \ac{pcrlb} in LOS condition as the \ac{gpr}-based likelihood interferes with the more precise physical model of the LOS component. It performs moderately during the \ac{olos} situation, losing the track in many realizations, showing increased performance when trained with the ``full" dataset compared to training with the ``sparse" dataset. The reduced performance of the method from \cite{Kram2022GPR} in \ac{olos} situation (w.r.t. the proposed algorithm) can be contributed to (i) the purely Gaussian model of the filter, which offers reduced numerical stability w.r.t. the heavy-tailed likelihood of probabilistic data association methods \cite{BarShalom1995} and (ii) the ``anomaly detection" method, which showed a high number of false alarms and missed detections in the investigated experiment. 
The method from \cite{WymeerschIEEE2012}, when trained with the ``reduced" dataset, shows significantly reduced performance even when the LOS to all anchors is available. It looses the track in many realizations starting at time $n=74$ when the LOS to anchor $1$ becomes unavailable. When we use the ``overfitted" dataset, it manages to perform robustly (i.e., no outliers), confirming its fundamental functionality. However, the \ac{rmse} in LOS condition is still significantly reduced. The conventional multi-sensor \ac{pda}, which does not facilitate the information contained in \acp{mpc}, performs well in LOS condition, but loses track in many realizations during \ac{olos} and follows wrong modes.

\section{Conclusion}\label{sec:conclusion}

 \acresetall
We have presented a neural-enhanced \ac{spa} that sequentially estimates the position of a mobile agent using radio signal measurements of multiple anchors by utilizing a hybrid probabilistic model that consists of physics-based and data-driven measurement models embedded in a joint Bayesian framework.
We analyzed the performance of the proposed algorithm using numerical simulation in a challenging scenario involving simultaneous obstructed line-of-sight (OLOS)\acused{olos} to all anchors. We demonstrated that our algorithm outperforms state-of-the-art methods for robust positioning and tracking, while consistently attaining the \ac{pcrlb} (i.e., no lost tracks) even with training data limited to local regions by fusing the information contained in the physics-based and data-driven measurement models.
Possible directions for future research include investigating alternative, uncertainty-aware regression methods for the data-driven measurement model to replace the %
\ac{gpr}-based \acf{lhf}. These methods should address the challenge of generalization to environments, which are not covered by training data, using concepts such as transfer learning, and should provide
 an efficient prediction step whose computational complexity does not directly depend on the size of the training data, such as Bayesian neural networks\cite{MacKay1992}. 

\vspace{-1mm}
 \section*{Acknowledgement}
\vspace{-1mm}
The authors thank Dr. Alexander Fuchs and Tobias Gailhofer for valuable discussions %
 during the course of this work.

 \acrodef{mimo}[MIMO]{multiple input multiple output}
 \acrodef{awgn}[AWGN]{additive white Gaussian noise}
 \acrodef{bw}[BW]{bandwidth}
 \acrodef{blt}[BLT]{bluetooth}
 \acrodef{cdf}[CDF]{cumulative distribution function}
 \acrodef{crlb}[CRLB]{Cram\'er-Rao lower bound}
 \acrodef{dmc}[DMC]{dense multipath component}
 \acrodef{dut}[DUT]{device under test}
 \acrodef{eirp}[EIRP]{equivalent isotropic radiated power}
 \acrodefplural{esl}[ESLs]{electronic shelf labels} 
 \acrodef{los}[LOS]{line-of-sight}
 \acrodef{mf}[MF]{matched filter}
 \acrodef{ml}[ML]{maximum likelihood}
 \acrodef{mpc}[MPC]{multipath component}
 \acrodef{nlos}[NLOS]{non-\ac{los}}
 \acrodef{pcb}[PCB]{printed circuit board}
 \acrodef{pdf}[PDF]{probability density function}
 \acrodef{reb}[REB]{ranging error bound}
 \acrodef{rss}[RSS]{received signal strength}
 \acrodef{smc}[SMC]{specular multipath component}
 \acrodef{snr}[SNR]{signal-to-noise-ratio}
 \acrodef{sinr}[SINR]{signal-to-interference-plus-noise-ratio}
 \acrodef{tdoa}[TDOA]{time difference of arrival}
 \acrodef{tka}[TKA]{trusted keyless access}
 \acrodef{toa}[TOA]{time-of-arrival}
 \acrodef{aoa}[AOA]{angle-of-arrival}
 \acrodef{uwb}[UWB]{ultra wide band}
 \acrodef{mie}[MIE]{method of interval estimation}
 \acrodef{mc}[MC]{Monte Carlo}
 \acrodef{mse}[MSE]{mean squared error}
 \acrodef{ci}[CI]{confidence interval}
 \acrodef{cl}[CL]{confidence level}
 \acrodef{pdp}[PDP]{power delay profile}
 \acrodef{dps}[DPS]{delay power spectrum}
 \acrodef{dm}[DM]{dense multipath}
 \acrodef{nlike}[NLIKE]{normalized likelihood}
 \acrodef{zzb}[ZZB]{Ziv-Zakai bound}
 \acrodef{ut}[UT]{unscented transform}
 \acrodef{glrt}[GLRT]{generalized likelihood ratio test}
 \acrodef{mse}[MSE]{mean squared error}
 \acrodef{rmse}[RMSE]{root mean squared error}
 \acrodef{nnlike}[NNLIKE]{normalized noise-free likelihood}
 \acrodef{stdv}[STDV]{standard deviation}
 \acrodef{rv}[RV]{random variable}
 \acrodef{bp}[BP]{belief propagation}
 \acrodef{pda}[PDA]{probabilistic data association}
 \acrodef{mp}[MP]{multipath}
 \acrodef{pmf}[PMF]{probability mass function}
 \acrodef{pdaf}[PDAF]{probabilistic data association filter}
 \acrodef{pdaai}[AIPDA]{amplitude-information \ac{pda}}
 \acrodef{olos}[OLOS]{obstructed \ac{los}}
 \acrodef{spa}[SPA]{sum-product algorithm}
 \acrodef{mmse}[MMSE]{minimum mean-square error}
 \acrodef{lhf}[LHF]{likelihood function}
 \acrodef{fa}[FA]{false alarm}
 \acrodef{ceda}[CEDA]{channel estimation and detection algorithm} 
 \acrodef{pcrlb}[P-CRLB]{posterior Cram\'er-Rao lower bound}
 \acrodef{slam}[SLAM]{simultaneous localization and mapping}
 \acrodef{mpslam}[MP-SLAM]{multipath-based SLAM}
 \acrodef{va}[VA]{virtual anchor}
 \acrodef{dnr}[DNR]{dense-to-noise ratio}
 \acrodef{aednn}[AE-DNN]{autoencoder deep neural network}   
 \acrodef{gpr}[GPR]{Gaussian process regression}  
 \acrodef{ae}[AE]{autoencoder}

\renewcommand{\baselinestretch}{0.92}\small\normalsize 

\bibliographystyle{IEEEtran}
\bibliography{IEEEabrv,references}

\end{document}